\documentclass{nature}
\usepackage{epsfig}
\usepackage[10pt]{moresize}
\usepackage{amssymb}
\usepackage{comment}
\usepackage{amsmath}

\newcommand{\ket}[1]{| #1 \rangle}
\newcommand{\bra}[1]{\langle #1 |}

\newcommand{\beq}{\begin{eqnarray}}
\newcommand{\eeq}{\end{eqnarray}}

\newcommand{\eq}[1]{Eq.~(\ref{#1})}

\begin{document}

\title{Delocalized single-photon Dicke states and the Leggett-Garg inequality in solid state systems}
\author{Guang-Yin Chen$^{1,2}$, Neill Lambert$^2$, Che-Ming Li$^{1}$, Yueh-Nan
Chen$^{1\star}$, and Franco Nori$^{2,4}$} \maketitle

\begin{affiliations}
\item
Department of Physics and
National Center for Theoretical Sciences, National Cheng-Kung
University, Tainan 701, Taiwan
\item
Advanced Science Institute,
RIKEN, Wako-shi, Saitama 351-0198, Japan
\item
Department of Engineering Science, National Cheng-Kung University,
Tainan City 701, Taiwan
\item
Physics Department, University of
Michigan, Ann Arbor, MI 48109-1040, USA
\\$^\star$e-mail:yuehnan@mail.ncku.edu.tw
\end{affiliations}

\begin{abstract}
We show how to realize a single-photon Dicke state in a
large one-dimensional array of two-level systems, and discuss how
to test its quantum properties. Realization of single-photon
Dicke states relies on the cooperative nature of the interaction
between a field reservoir and an  array of two-level-emitters. The
resulting dynamics of the delocalized state can display Rabi-like
oscillations when the number of two-level emitters exceeds several
hundred. In this case the large array of emitters is essentially
behaving like a ``mirror-less cavity''. We outline how this might
be realized using a multiple-quantum-well structure and discuss
how the quantum nature of these oscillations could be tested with
the Leggett-Garg inequality and its extensions.
\end{abstract}

When an ensemble of atoms interacts with a common radiation field
each atom can no longer be regarded as an individual radiation
source but the whole ensemble of atoms can be regarded as a
macroscopic dipole moment\cite{Lvo, Inouye}. This collective
behavior leads to cooperative radiation, i.e. the so-called
superradiance introduced by Dicke in 1954. Superradiance, and its
extended effects, has also been observed in solid state systems
such as quantum dots\cite{SRQD}, quantum wells\cite{SRQW}, and
coupled cavities\cite{zhou}.  This effect is generally
characterized by an enhanced emission intensity that scales as the
square of the number of atoms.

Recently, a particularly interesting consequence of this
cooperative interaction was discussed by Svidzinsky \textit{et al}
\cite{scullyPRA,newPRA,scullyPRL08}.  In their work they showed
that there could be cooperative delocalized effects even when just
a single photon is injected into a large cloud of atoms.  The
state that is created via this mechanism is a highly-entangled
Dicke state\cite{wstatepaper}. An interesting open question is if
such a state can be realized and manipulated in a solid-state
environment.

To answer this question we analyze what happens when a
single-photon is injected into a large \textit{one-dimensional}
array of two-level-emitters (TLE). We find that because of the
cooperative interaction between light and matter the structure
acts like an effective optical cavity without mirrors\cite{scullyPRL08}, and realizes a one-dimensional variation of
the Dicke-state discussed by Svidzinsky \textit{et al} \cite{scullyPRA,newPRA,scullyPRL08}. We show that the delocalized
state formed in this emitter-array can exhibit quantum behaviour
through the coherent oscillatory dynamics of the state. We discuss
how such a phenomenon might be realized in a multiple-quantum-well
(MQW) array and discuss physically-realistic parameters.  To show
how the quantum features of such an experiment might be verified,
we apply the Leggett-Garg (LG) inequality\cite{LG}, and a
Markovian extension\cite{NLPRL}, to examine the quantum coherence
of the delocalized state over the MQW structure. Finally, we
discuss two other potential candidates for experimental
realization.

\section*{Results}
We consider an array containing $N$ two-level emitters
coupled to a photonic reservoir. A photon with wavevector $k_0$
incident on the array, as shown in Fig.~1(a). If the $N$-TLE array
uniformly absorbs this incident photon (in practice, one can
detune the incident photon from resonance, such that TLEs are
equally likely to be excited\cite{scullyscience}), the $N$-TLE
can be in a collective excited state with one excitation
delocalized over the whole system.  Post-selecting this state
(since in the vast majority of cases the photon will not be
absorbed) results in the superposition state
\begin{equation}
|+\rangle_{\textbf{k}_0}=\frac{1}{\sqrt{N}}\sum_je^{ik_0z_j}|j\rangle
\end{equation}
of the exciton in this $N$-TLE structure, where $z_j$ is the position
of the $j$th TLE. The state
\begin{equation}
|j\rangle=|\textrm{g}_1,\textrm{g}_2,...,\textrm{g}_{j-1},\textrm{e}_j,\textrm{g}_{j+1},...,\textrm{g}_N\rangle
\end{equation}
describes the state with the $j$th TLE being in its
excited state. Including the coupling between the TLE array and the
1D radiation fields, the state vector of the total
system at time $t$ can be written as:
\begin{equation}
|\Psi(t)\rangle=b_+(t)|+\rangle_{\textbf{k}_0}|0\rangle+b_{\bot}(t)|\bot\rangle_{\textbf{k}_0}|0\rangle+\sum_{k_{z}}b_{k_{z}}(t)|\textrm{g}\rangle|1_{k_{z}}\rangle,
\label{ket}
\end{equation}
where $|0\rangle$ denotes the zero-photon state,
$|1_{k_{z}}\rangle$ denotes one photon in the $k_{z}$-mode, and
$|\textrm{g}\rangle$ is the TLE ground state. Note that
the superposition state $|+\rangle_{\textbf{k}_0}$ is a Dicke
state\cite{ scullyPRL08, scullyPRL06, scullylaser}, and
$|\bot\rangle_{\textbf{k}_0}$ is a summation over all other Dicke
states orthonormal to $|+\rangle_{\textbf{k}_0}$. The interaction
between the TLE array and radiation fields can then be described
by\cite{Lee, YN}
\begin{equation}
H_{\textrm{int}}=\sum_{k_{z}}\sum_{j=1}^{N}\hbar
g_{k_{z}}\{{\sigma^-_ja^{\dag}_{k_{z}}e^{[i(\omega_{k_{z}}-\omega_0)t-ik_{z}z_j]}}+\textrm{h.c.}\},
\label{H}
\end{equation}
where $\omega_{k_{z}}$ is the frequency of the
$k_z$-mode photon, $\omega_0$ is the excitation energy of the TLE, $\sigma^-_j$ is
the lowering operator for the $j$th TLE, $a^{\dag}_{k_{z}}$ is the creation operator for one photon in the
$k_z$-mode, and $g_{k_{z}}$ is the coupling strength between
TLE and the $k_z$-mode photon.

In the limit of $k_0L\gg1$ ($L$ is the total length of
the array), from the time-dependent Schr\"{o}dinger equation
\begin{equation}
i\hbar\frac{\partial}{\partial t}|\Psi(t)\rangle=H_\textrm{int}|\Psi(t)\rangle,
\label{sch}
\end{equation}
 the dynamical evolution of the Dicke state
$|+\rangle_{\textbf{k}_0}$ can be written as\cite{scullyPRL08}:
\begin{equation}
\dot{b}_+(t)=-\frac{1}{N}\int_0^t\!\!\!\!dt'\sum_{k_z}\sum_{i,j=1}^Ng_{k_z}^2[e^{i(\omega_{k_z}-\omega_0)(t'-t)}e^{i(k_z-k_0)(z_i-z_j)}]b_+(t').
\label{b}
\end{equation}
With the approximation $g_{k_z}^2\approx g_{k_0}^2$ and $\sum_{k_z}\rightarrow \textrm{L}_{\textrm{ph}}/(2\pi)\int dq$, Eq.~(\ref{b}) can be expressed as:
\begin{equation}
\dot{b}_+(t)=-\frac{1}{N}\frac{\textrm{L}_{\textrm{ph}}}{2\pi}g_{k_0}^2\int_0^t\!\!\!\!dt'\;b_+(t')\int_{-\infty}^{\infty}\!\!\!\!dq\;\{e^{ivq(t'-t)}\sum_{\xi=0}^N[(N-\xi)(e^{i\xi qh}+e^{-i\xi qh})]\},
\label{b2}
\end{equation}
where $\textrm{L}_{\textrm{ph}}$ is the quantization length of the radiation field, $v$ is the speed of light, and $\xi$ is a counting index, since the value of $(z_i-z_j)$ can range between $-Nh$ and $Nh$. The dynamical evolution of the Dicke state $|+\rangle_{k_0}$ can thus be obtained by solving Eq.~(\ref{b2}).

For the array containing $N$ TLEs, the dynamical evolution of the
state $|+\rangle_{\textbf{k}_0}$ can be enhanced by the
superradiant effect,
$\Gamma_{\textrm{array}}=N~\Gamma_{\textrm{TLE}}$, as shown in the
red dashed and blue dotted curves shown in Fig.~1(b). For an
extremely large array ($L\gg\lambda$, where $\lambda$ is the
wavelength of the emitted photon), the probability to be absorbed across
the whole sample is made uniform by sufficiently detuning the
incident photon energy from that of the TLEs\cite{scullyscience}. As mentioned earlier
this means that the majority of photons pass through unabsorbed.
Later we will discuss how the absorbtion event can be signalled by
a two-photon correlation when this scheme is realized by an array
of quantum wells.

The solid curve in Fig.~1(c) represents Rabi-like oscillations
together with an exponential decay. The enhanced decay rate proportional to $N$ is a quantum
effect, but may also be described in a semi-classical way by
regarding the $N$ TLEs as $N$ classical harmonic
oscillators\cite{scullyPRA}.
For $N\gg1$, the summation $\sum_{i,j=1}^N$ in Eq.~(\ref{b}) can
be replaced by the integration $(N/L)^2\int\!dz\int\!dz'$, showing
that the effective coupling strength $g$ between the state
$|+\rangle_{\textbf{k}_0}$ and the field is
$g=\sqrt{N}g_{\textbf{k}_0}$. The period of oscillations is
therefore enhanced by a factor $\sqrt{N}$ compared to the bare
exciton-photon coupling.

\subsection{Effective two-level system}

To illustrate that the Rabi-like oscillation is mathematically
equivalent to an effective quantum coherent oscillations between two states (e.g., a spin
or a single excitation cavity-QED system), we transform the
Eq.~(\ref{b2}) into the energy
representation via $\tilde{b}_+(E)=\int_0^\infty b_+(t)e^{iEt}dt$, and obtain\cite{Gurvitz}:
\begin{equation}
\left\{E+\frac{1}{N}\frac{\textrm{L}_{\textrm{ph}}}{2\pi}\;g_{k_0}^2\int_{-\infty}^{\infty}~\!\!\!dq~\frac{\sum_{\xi=0}^N[(N-\xi)2\cos(\xi qh)]}{E-vq}\right\}b_+(E)=-i.
\label{dos}
\end{equation}
Equation~(\ref{dos}) thus indicates that the density of states (DOS) $D(q)$ of the radiation field in the TLE array,
\begin{equation}
D(q)\propto\sum_{\xi=0}^N[(N-\xi)\cos(\xi
qh)],
\end{equation}
where $q\equiv k_z-k_0$, $\xi$ is a counting index, and $h$
denotes the separation between each period. The insets in Fig.~1(b) and 1(c) show the
DOS for TLE array containing different number of emitters. As can be seen, when
increasing the number of periods $N$, the line-shape of $D(q)$
(black solid curve in the inset of Fig.~1(c)) becomes Lorentzian-like. Therefore, the TLE array coupled to radiation fields can be interpreted as a Dicke state $|+\rangle_{\textbf{k}_0}$
coupled to a Lorentzian-like continuum, as shown in Fig.~2(a). Following the study by Elattari and
Gurvitz\cite{Gurvitz}, for large $N$, our system can be mapped to
the Dicke state $|+\rangle_{\textbf{k}_0}$ coherently coupled to a resonant
state $|k_0\rangle$ with a Markovian dissipation as depicted in
Fig.~2(b). The remaining part of the DOS which does not fit the
Lorenzian distribution can be treated as an effective excitonic
polarization decay.

\subsection{Realization with multiple-quantum-wells}

To show that this effect can be realized in a solid-state
environment we consider in detail how to use a
multiple-quantum-well (MQW) structure as the two-level-emitter
array. In such a MQW structure, each single quantum well can be
regarded as a two-level emitter. The quantum-well exciton will be
confined in the growth direction (chosen to be \textit{z}-axis)
and free to move in the \textit{x}-\textit{y}-plane. Due to the
relaxation of momentum conservation in the \textit{z}-axis, the
coupling between the photon fields and the quantum wells is
one-dimensional. Therefore, if we assume a incident photon with
wavevector $k_0$ on the MQW along the \textit{z}-axis, the
interacting Hamiltonian can be written exactly the same as the
form in Eq.~(\ref{H}). Furthermore, quantum wells have the
remarkable advantage that  the phase factor $ik_0z_j$ in
$|+\rangle_{k_0}$ can be fixed during the quantum-well growth
process, and since the photon fields travel in MQW only along the
\textit{z}-axis, a one-dimensional waveguide is not required.

To elaborate on the physical parameters necessary to realize the
single-photon Dicke state we assume a MQW structure with a period
of 400 nm, where each quantum well consists of one GaAs layer of
thickness 5 nm (sandwiched between two AlGaAs slabs). The exciton
energy $\hbar\omega_0$ of a single quantum well can take the
value\cite{pin} we utilized in Fig.~1 (i.e., 1.514 eV), such that
the resonant photon wavelength $\lambda=2\pi c/\omega_0\approx$
820 nm.  To realize the Dicke state at all we already demanded
that the photon be off-resonance with the array. In principle the
on-resonance regime can be reached by tuning the quantum well
array energies after the Dicke state has been realized. To
identify when the state has been created a pair of identical
photons with wavevector $k_0$ are produced by the two-photon
down-conversion crystal,  as shown in Fig.~3. One of the photons
is directed to the detector-1 (D1) and the other is along the
growth direction of the MQW. The distance between the crystal and
D1 is arranged to be the same as that between the crystal and the
MQW. Once there is a click in D1, there should be one photon
simultaneously sent into the MQW. The photon incident on the MQW
generally passes through the MQW and registers a count in
detector-2 (D2), but it could also excite one of the multiple
quantum wells and form a delocalized exciton. The presence of a
count in D1 and the absence of a count in D2 therefore tells us
that the MQW has been prepared in the superposition state
$|+\rangle_{k_0}$. Since the interaction between the photon fields
and the MQW structure is identical with Eq.~(\ref{H}), the exciton
dynamics of the $|+\rangle_{k_0}$ and the density of states of the
photon fields in MQW can show the same behaviors as those in
Fig.~1(b) and (c) (here one unit of time is $10$ picosecond) when
the MQW contains corresponding number $N$ of the quantum wells.

For a MQW structure containing a large number of quantum wells
(i.e., $N\geq200$), the dynamical evolution of the superposition
state $|+\rangle_{k_0}$ shows Rabi-like oscillations. However, one
should note that the Rabi-like oscillations here are different
from the Rabi oscillations reported in secondary emission
spectra\cite{koch, kavokin} of excitons in the MQW structures.
The secondary emission occurs when the MQW is illuminated by
coherent light, and emission occurs in a direction different from
the excitation direction. However, in our system, the incident
excitation is a {\em single photon}, and the detector-2 (see
Fig.~3) receiving the emitted photon is positioned along the
excitation direction. Furthermore, the MQW system we consider is
Bragg-arranged (i.e., the inter-well spacing equals half the
wavelength of light at the exciton frequency), for which the Rabi
oscillations in secondary emission cannot appear\cite{kavokin}.
Therefore, the Rabi-like oscillations in Fig.~1(c) are different
from those in secondary emission but are a result of the coherent
oscillations between the delocalized exciton state
$|+\rangle_{\textbf{k}_0}$ and the resonant photon state
$|k_0\rangle$.

\subsection{The Leggett-Garg Inequality}

While we have argued that the oscillations one would observe in
this large mirror-less cavity are quantum-mechanical in nature
(akin to vacuum Rabi splitting), there is still some ambiguity. In
the earlier work of Svidzinsky \textit{et al} \cite{scullyPRA}
they employ a semi-classical explanation of a similar phenomena.
Thus the question remains open as to whether $\ket{k_0}$ can truly
be considered a single resonant state with neglible phase
decoherence, and whether the Dicke state retains its
long-range spatial coherent nature on a sufficient time-scale.
There may be alternative classical and semi-classical explanations
of the oscillations one may
see in experiment.

Similar problems were overcome in the field of cavity and
circuit-QED by observation of other quantum features (e.g., the
scaling of the energy spectrum\cite{CircuitQED}). However, in
quantum wells we are restricted to certain types of measurements.
 Recently, great advances have been made in measuring the excitonic
states in quantum wells via four-wave mixing
techniques\cite{Patton}. We can also, in principle, make
measurements on the emitted photons (e.g., as in Fig.~3). As a
first test one could measure the emitted photons at D2 to verify
the single-particle nature of the dynamics with a simple violation
of the Cauchy-Schwarz inequality\cite{QuantumNoise}, or
observation of anti-bunching. We will not go into detail on this
here, but essentially it corresponds to the detection of only
single-photons. In other words, after we detect the single photon
at D2, no more measurements will occur until another photon is
injected into the sample.  This is a trivial application of the
Cauchy-Schwartz inequality, but indicates that we are operating in
the single-excitation limit.

In order to verify the quantum coherence of the delocalized state
in the MQW rigorously one could apply a test like the Leggett-Garg
(LG) inequality\cite{LG}.  The LG inequality depends on the fact
that at a macroscopic level several assumptions about our
observations of classical reality can be made: realism, locality,
and the possibility of non-invasive measurement. In 1985, Leggett
and Garg derived their inequality\cite{LG} to test the first and
last assumptions, which when combined they called ``macroscopic
realism''. The experimental violations of this inequality in a
``macroscopic'' superconducting circuit\cite{Laloy}, polarized
photon state\cite{PNAS, PRL, SR}, electron-nuclear spin
pairs\cite{natcomm, PRL_spin}, have recently been seen.

Given a dichotomic observable $Q(t)$, which is bound by
$|Q(t)|\leq1$, the Leggett-Garg inequality is:
\begin{equation}
|L_Q(t)|\equiv|\langle Q(t_1)Q\rangle+\langle Q(t_1+t_2)Q(t_1)\rangle-\langle Q(t_1+t_2)Q\rangle|\leq1,
\label{LG}
\end{equation}
where $Q\equiv Q(t=0)$, and $t_1<t_2$.  A violation of this
inequality suggests either the assumption of realism or of
non-invasive measurements is being broken.

To apply this to the system we have been discussing we must
formalize further how, for large $N$, the MQW system can be mapped
to an effective two-level system [as shown in Fig.~2(c)]. The
dynamics of this effective model can be described by a Markovian
master equation:
\begin{equation}
\dot{\rho}=\mathcal{L}[\rho]=\frac{1}{i\hbar}[\widetilde{H}_{\textrm{eff}},\rho]+\Sigma[\rho],
\label{Liou}
\end{equation}
where
\begin{equation}
\begin{array}{ll}
                \widetilde{H}_{\textrm{eff}}=&\hbar g(\sigma^{-}+\sigma^{+})\\
                \Sigma[\rho]=&\kappa(s\rho s^{\dag}-\frac{1}{2}s^{\dag}s\rho-\frac{1}{2}\rho s^{\dag}s)+\gamma(r\rho r^{\dagger}-\frac{1}{2}r^{\dagger}r\rho-\frac{1}{2}\rho r^{\dagger}r).
                \end{array}
\label{master}
\end{equation}

Here, $\mathcal{L}$ is the Liouvillian of the system,
$\widetilde{H}_{\textrm{eff}}$ is the coherent interaction in this
effective cavity-QED system,
$\sigma^{-}=|k_0\rangle_{\textbf{k}_0}\langle+|$
($\sigma^{+}=|+\rangle_{\textbf{k}_0}\langle k_0|$) denotes the
lowering (raising) operator for the Dicke state
$|+\rangle_{\textbf{k}_0}$, and $g=\sqrt{N}g_{\textbf{k}_0}$. The
state $|\textrm{vac}\rangle$ is the vacuum state which in the full
basis is $|\textrm{g}\rangle\otimes |0\rangle$, i.e. no excitation
in the Dicke state or in the resonant state $k_0$. In the
self-energy $\Sigma[\rho]$, the $s=| \mathrm{vac}\rangle \langle
k_0|$ operators describe the loss of the photon from the MQW
system with rate $\kappa$, and the $r =
|\mathrm{vac}\rangle_{\textbf{k}_0}\langle+|$ operators describe
the loss of excitonic polarization with rate ${\gamma}$. With this
master equation, in Fig.~4(a) we plot $|L_Q(t)|$ using the
observable\cite{PRB_NL},
\begin{equation}
Q=|k_0\rangle\langle k_0|-|+\rangle_{\textbf{k}_0~\textbf{k}_0}\langle+|-|\mathrm{vac}\rangle\langle\mathrm{vac}|.
\end{equation}
Considerable violations ($>1$) of the LG inequality
[Eq.~(\ref{LG})] appear in the region above the blue dashed line in
Fig.~4(a). The violations resulting from the quantum oscillations
between the states $|+\rangle_{\textbf{k}_0}$ and $|k_0\rangle$
indicate the quantum coherence of the delocalized state in the MQW
structure.

A direct application of this inequality to the example of a
quantum well array seems extremely challenging because the
measurement of a photon leaving the system, and the four-wave
mixing measurements of the excitonic states
\cite{Patton,Schulzgen}, are fundamentally invasive. To test the
inequality unambiguously would require a fast projective (quantum
non-demolition) measurement of the single photon state
$\ket{k_0}$, or the Dicke state $\ket{+}_{\textbf{k}_0}$. Such
measurements are now in principle possible in
optical\cite{Gleyzes, Khalili} and microwave \cite{Johnson, NLPRA}
cavities, but not in the effective cavity we describe here.

Some progress can be made by making further assumptions. It was
shown by Huelga et al\cite{huelga2, huelga3, huelga1} and
others\cite{NLPRL, NLPRA} that the assumption of Markovian
dynamics eliminates the need to assume non-invasive measurement if
we can reliably prepare the system in a desired state (then the
invasive nature of the second measurement, e.g., because of the
destruction of the photon, does not affect the inequality). Under
this Markovian assumption the inequality can be written in terms
of population measurements of the state we wish to measure (which
in general we describe as a single-state projective operator
$Q=\ket{q}\bra{q}$, for some measurable state of the system
$\ket{q}$),
\begin{equation}
|L_{P_{Q}}(t)|\equiv|2\langle P_{Q}(t)P_{Q}\rangle-\langle
P_{Q}(2t)P_{Q}\rangle|\leq \langle P_{Q}\rangle, \label{L1}
\end{equation}
where $\langle P_{Q}\rangle$ is the expectation value of the
zero-time population $P_{Q}\equiv P_{Q}(t=0)$, and $\langle
P_{Q}(t)P_{Q}\rangle$ is the two-time correlation function. Note
that if the zero-time state is the steady state then this is
equivalent to the original\cite{LG} LG inequality, but again
demands non-invasive measurements. If the zero-time state is not
the steady state, but some prepared state e.g. $\rho(0)=Q$,
$P_Q(0)=1$, then a violation of this variant of the Leggett-Garg
inequality indicates behaviour only beyond a classical Markovian
regime, i.e. a strong indication of the quantumness of this
delocalized state, though not irrefutable proof.  We now consider
the above inequality in two different regimes.

\subsection{Initial Dicke state:  Markovian test}

If we can deterministically prepare the state $\ket{+}$ (dropping
the $k_0$ subscript for brevity) as described in Fig.~3, we can
construct the inequality (\eq{L1}) with $\ket{q}=\ket{+}$ by
preparing that state so $P_{+}(0)=1$, and then (invasively)
measuring the state of the quantum wells at time $t$ later (see
below). This is then equivalent to the test to eliminate purely
Markovian dynamics\cite{huelga2, huelga3, huelga1}.
 In general such a measurement will be invasive (and can generally
be described by some positive operator valued measurement (POVM)),
but since we are not concerned with events after the second
measurement, we can just assume that it is proportional to the
probability of obtaining the Dicke state $\ket{+}$ at that time.
In other words, we can assume the second measurement is just a
normal projective measurement, $P_{+}\equiv|+\rangle\langle +|$.

The correlation function $\langle P_{+}(t)P_{+}\rangle$, where
$P_{+}(0)=1$, can be calculated from
\begin{equation}
\langle
P_{+}(t)P_{+}\rangle=\textrm{Tr}[P_{+}\exp(\mathcal{L}t)\ket{+}\bra{+}]
\end{equation}
In Fig.~4(b), we plot $|L_{P_{+}}(t)|$ as a function of
time (solid black curve). The behavior is oscillatory but damped
due to the couplings to the Markovian photon dissipation and the
excitonic polarization decay. A considerable violation $(>1)$ of
the inequality of Eq.~(\ref{L1}) appears in the region above the blue
dashed line in Fig.~4(b). The violation there comes from the
coherent oscillations between the states $|+\rangle$ and
$|k_0\rangle$, and is beyond the classical Markovian description.

The Dicke state $\ket{+}$ describes a particular
coherent superposition of a single excitation across all $N$
quantum wells. It has been shown that four-wave mixing and pump
probe techniques \cite{Patton,Schulzgen} can be used to measure
the state of multiple excitations across multiple wells.  Thus it
seems feasible that such an experiment can be used to determine
the excitation density.  If we assume that only the $\ket{+}$
plays a role here, and the other Dicke states ($|\bot\rangle$) are
unoccupied, then this is sufficient for our purposes, as it will
tell us if the array contains a single excitation or not. However,
whether the $\ket{+}$ can be in general distinguished from the
other Dicke states with such a measurement is an interesting open
problem, and requires further study.

\subsection{Initial photonic state:  Markovian test}

Similarly, if we could deterministically prepare the state
$\ket{k_0}$, we could construct the inequality (\eq{L1}, with
$\ket{q}=\ket{k_0}$) by preparing that state (so $P_{k_0}(0)=1)$,
and then measuring when a single photon is detected at detector
D2. The second measurement needed to construct the correlation
functions in \eq{L1} is then simply given by the superoperator
\begin{equation}
\mathcal{J}(\rho)=\kappa|\mathrm{vac}\rangle_{k_0}\langle
k_0|\rho|k_0\rangle_{k_0}\langle \mathrm{vac}|,
\end{equation}
where $\ket{\mathrm{vac}}$ is the vacuum state. Again, we can
assume the second measurement is just a normal projective
measurement (after rescaling by $\kappa$),
$P_{k_0}\equiv|k_0\rangle\langle k_0|$. Thus, while the photon
measurement is much simpler than the quantum well one described
earlier, in our scheme it is not clear if we can determinstically
know when $\ket{k_0}$ is created in the same way that $\ket{+}$
is, as $\ket{k_0}$ is an effective state of the field modes. In
Fig.~4(b), we plot $|L_{P_{k_0}}(t)|$ as a function of time (dashed
red curve). Again a considerable violation $(>1)$ of the
inequality of Eq.~(\ref{L1}) appears, and indicates behavior beyond
the classical Markovian description.

Of course, ultimately we cannot distinguish classical
non-markovian dynamics from quantum dynamics with this method,
though certain complex Markovian systems can produce non-monotonic
and complex behavior\cite{NLPRL} which it is important to
eliminate. To really show that the large array of quantum wells is
behaving like a cavity without a mirror and exhibiting quantum
Rabi oscillations more work needs to be done on full state
tomography techniques and precise measurements of excitonic
states, so that either the full Leggett-Garg inequality, or some
other test, can be investigated.

\section*{Discussion}
In summary, we investigated the dynamical evolution of the
delocalized state of a two-level-emitter array state. When the
array contains a large number of emitters, the dynamical evolution
shows Rabi-like oscillatory behavior. By showing that the DOS of
the radiation field in the TLE array is Lorentzian-like, the whole
system can be mapped to an effective two-level system (e.g., like
a single excitation cavity-QED system). For the physical
implementation we suggested a multiple-quantum-well structure and
discussed relevant parameters. We also applied the original
Leggett-Garg inequality, and a Markovian variation of it, to
examine the quantum coherence of the MQW structure.

In addition to the MQW structure, there are other
experimentally-accessible systems that can mediate one-dimensional
coupling between two-level emitters and the photon fields. Below
we provide two potential candidates:

(I.) \textit{Metal nanowire}:~$N$ two-level quantum dots
positioned near a metal nanowire\cite{PRB_GY} as shown in
Fig.~5(a). Due to the quantum confinement, the surface plasmons propagate along the axis direction on the surface of the
nanowire. The coupling between quantum dots and the surface plasmons enable the incident
surface plasmons to excite one of the $N$ quantum dots and the delocalized exciton over
the $N$ dots can then be formed.

(II.) \textit{Superconducting transmission line}:~A
superconducting transmission line resonator coupled to $N$
dc-SQUID-based charge qubits\cite{zhou} as depicted in Fig.~5(b).
With proper gate voltage, the Cooper-pair box formed by the dc
SQUID with two Josephson junctions can behave like a two-level
system (charge qubit). The incident photon propagating in the
one-dimensional transmission line would excite one of the charge
qubits and form the delocalized state over the $N$ charge qubits.
Recent progress in generating and measuring single microwave
photons\cite{Mpho1,Mpho2,Mpho3,Mpho4,Mpho5} may make the generation and detection of the single-photon
Dicke state feasible in the near future.

\section*{Methods}
\noindent \textbf{Dicke states.} The state
$|\bot\rangle_{k_0}|0\rangle$ in Eq.~(\ref{ket}) denotes a
collection of single-excitation Dicke states besides
$|+\rangle_{k_0}$. The set of Dicke states are listed in the Table
I:

\begin{tabular}{lll}\hline\hline
$|+\rangle_{k_0}=\frac{1}{\sqrt{N}}\sum_je^{ik_0z_j}|j\rangle$\\
$|1\rangle_{k_0}=\frac{1}{\sqrt{2}}(e^{ik_0z_1}|1\rangle-e^{ik_0z_2}|2\rangle)$\\
$|2\rangle_{k_0}=\frac{1}{\sqrt{6}}(e^{ik_0z_1}|1\rangle+e^{ik_0z_2}|2\rangle-2e^{ik_0z_3}|3\rangle)$\\
\vdots\\
$|N-1\rangle_{k_0}=\frac{1}{\sqrt{N(N-1)}}[e^{ik_0z_1}|1\rangle+e^{ik_0z_2}|2\rangle+\ldots+e^{ik_0z_{N-1}}|N-1\rangle-(N-1)e^{ik_0z_{N}}|N\rangle]$\\ \hline\hline
\end{tabular}

\begin{addendum}

\item [Acknowledgement]

This work is supported partially by the National Science Council,
Taiwan, under the grant number NSC 98-2112-M-006-002-MY3 and NSC 100-2112-M-006-017. N.L. is supported by RIKEN's FPR scheme. F.N.
acknowledges partial support from the Laboratory of Physical
Sciences, National Security Agency, Army Research Office, Defense
Advanced Research Projects Agency, Air Force Office of Scientific
Research, National Science Foundation Grant No. 0726909, JSPS-RFBR
Contract No. 09-02-92114, Grant-in-Aid for Scientific Research (S),
MEXT Kakenhi on Quantum Cybernetics, and Funding Program for
Innovative R\&D on S\&T (FIRST).

\item [Author Contributions] GYC carried out all calculations under the guidance of NL and YNC. CML and FN attended the discussions.
All authors contributed to the interpretation of the work and the writing of the manuscript.

\item [Competing Interests]
The authors declare that they have no competing financial interests.

\item [Correspondence]
Correspondence and requests for materials should be addressed to Y.N.C.
\end{addendum}

\clearpage


\textbf{Table I} The set of all Dicke
states\cite{scullyPRL08}. Here,
$|j\rangle=|\textrm{g}_1,\textrm{g}_2,...,\textrm{g}_{j-1},\textrm{e}_j,\textrm{g}_{j+1},...,\textrm{g}_N\rangle$
describes the state with the $j$th two-level emitter in its excited
state.
\bigskip

\textbf{Figure 1: Dynamical evolution of the Dicke state and the
density of states of the radiation field in the two-level-emitter
array.} (a) The schematic diagram of the two-level-emitter array.
The array contains $N$ two-level emitters coupled to the
one-dimensional photon reservoir. With proper excitation energy,
the incident photon can excite one of the $N$ two-level emitters,
and the Dicke state can be formed. The dynamical evolutions of the
Dicke state $|+\rangle_{k_0}$ for the TLE array containing (b) 20
(red dashed), 60 (blue dotted), and (c) 200 (black-solid)
two-level emitters. These evolutions are obtained by solving the
time-dependent Schr\"{o}dinger equation
[Eq.~(\ref{sch})$\sim$(\ref{b2}) ] in the limit of $k_0L\gg1$. The
period of the oscillations for the black solid curve is 0.54 time
units. Here, the unit of time is normalized by the spontaneous
decay rate $\Gamma_{\textrm{TLE}}$ of a single two-level emitter. The insets show that when increasing the number of
periods $N$, the normalized density of states of the radiation
field in the TLE array containing 20 (red dashed), 60 (blue
dotted) [the inset in (b)], and 300 (black solid) [the inset in
(c)] two-level emitters. The green dashed-dotted curve of the
inset in (c) is a Lorentzian fit for $N$=200.
\bigskip

\textbf{Figure 2: The correspondence of two-level-emitter array to
other systems.} (a) The two-level-emitter array coupled to the
radiation field can be interpreted as the Dicke state
$|+\rangle_{k_0}$ coupled to a Lorentzian-like continuum spectrum
if $N$ is large enough. (b) The system can be further mapped to a
Dicke state coherently coupled to a resonant state $|k_0\rangle$
with a Markovian dissipation. The coupling strength $g$ between
$|+\rangle_{k_0}$ and $|k_0\rangle$ is $g=\sqrt{N}g_{k_0}$.
\bigskip

\textbf{Figure 3: Multiple-quantum-well structure.} A schematic
diagram of the GaAs/AlGaAs MQW structure. We assume that the MQW
structure is grown along the \textit{z}-axis, with a period of 400
nm, and each quantum well consists of one GaAs layer of thickness
5 nm (sandwiched between two AlGaAs slabs). The exciton energy
$\hbar\omega_0$ of a single quantum well is set to be\cite{pin}
1.514 eV, such that the resonant photon wavelength $\lambda=2\pi
c/\omega_0\approx$ 820 nm. A pair of identical photons with
wavevector $k_0$ could be produced by a two-photon down-conversion
crystal. One of the photons is directed to the detector-1 (D1) and
the other is along the growth direction of the MQW.
\bigskip

\textbf{Figure 4: Violation of the LG inequality and its
extensions.} (a) The original LG inequality for $|L_Q(t)|$
[Eq.~(\ref{LG})] as a function of time. The region above the blue
dashed line indicates the violation regime. (b) The inequality
   $|L_{P_{Q}}(t)|$ [Eq.~(\ref{L1})] as a
   function of time for  the state
   $\ket{q}=|k_0\rangle$ (red dashed curve) and  $\ket{q}=|+\rangle_{\textbf{k}_0}$ (black solid curve)  in a MQW system containing 200 periods.  The region above the blue dashed line indicates
the violation regime.  In plotting both figures, the coupling
constant $g=8.3$ meV, between $|+\rangle_{\textbf{k}_0}$ and
$|k_0\rangle$, is determined from the period of the Rabi-like
oscillations in Fig.~1(c). The photon loss $\kappa=3.3$ meV is
obtained from the width of the Lorentzian fitting (the green
dashed-dotted curve in the inset of Fig.~1(c). Here we have set
the excitonic polarization decay rate $\gamma$ as the spontaneous
emission rate of the general GaAs/AlGaAs quantum well
$\gamma=\Gamma_{QW}=100$ (1/ns).
\bigskip

\textbf{Figure 5: Experimentally-accessible systems.} Schematics
of two alternative experimentally-accessible systems which
could realize single-photon Dicke states: (a) $N$ two-level quantum
dots coupled to metal nanowire surface plasmons; and (b) $N$
dc-SQUID-based charge qubits coupled to a one-dimensional
transmission line. A Cooper-pair box formed by a DC SQUID with two
Josephson junctions can act like a two-level system by properly
tuning the gate voltage. The incident surface plasmon (photon in
transmission line) can excite one of the $N$ quantum dots (charge
qubit), the excited quantum dot (charge qubit) can re-emit a surface plasmon (photon) which would be absorbed by
another quantum dot (charge qubit), and so on. A delocalized
state over the quantum-dot (charge-qubit) array can therefore be
formed. \clearpage

\clearpage
\begin{figure}
\begin{center}
\epsfig{file=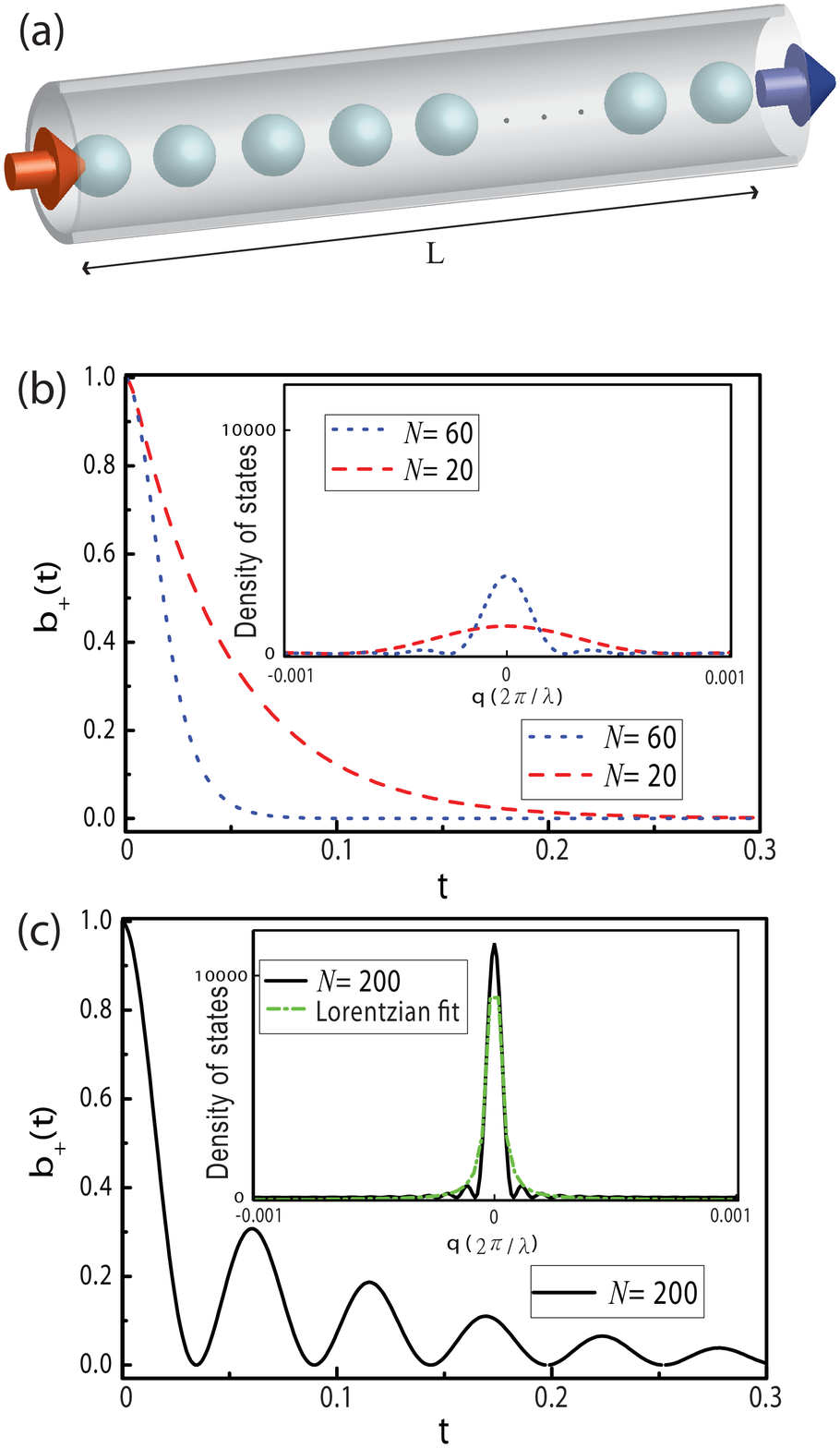,width=12cm}
\end{center}\caption{}
\label{fig1-scheme}
\end{figure}

\clearpage
\begin{figure}
\begin{center}
\epsfig{file=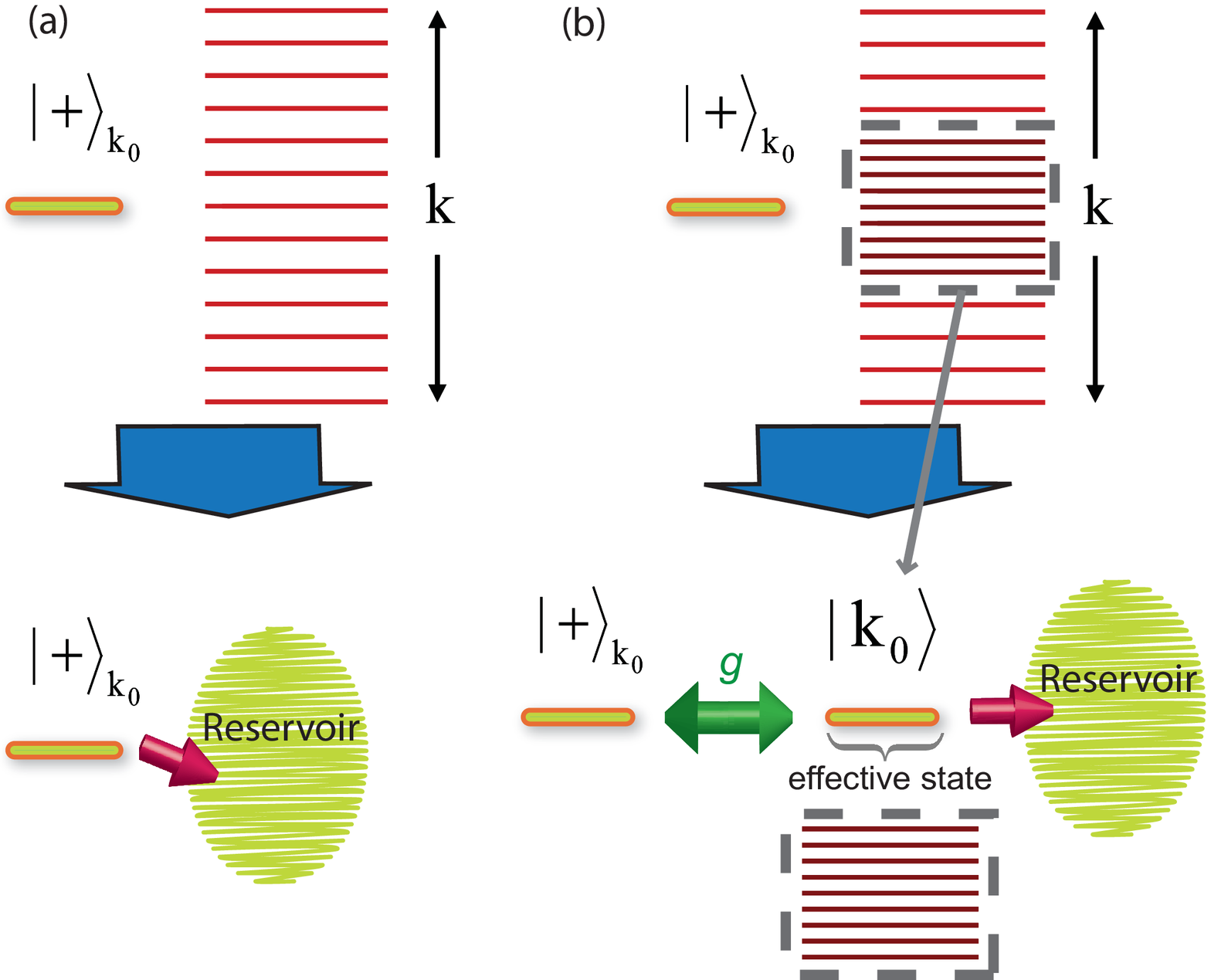,width=14cm}
\end{center}\caption{}
\label{fig1-scheme}
\end{figure}
\clearpage

\begin{figure}
\begin{center}
\epsfig{file=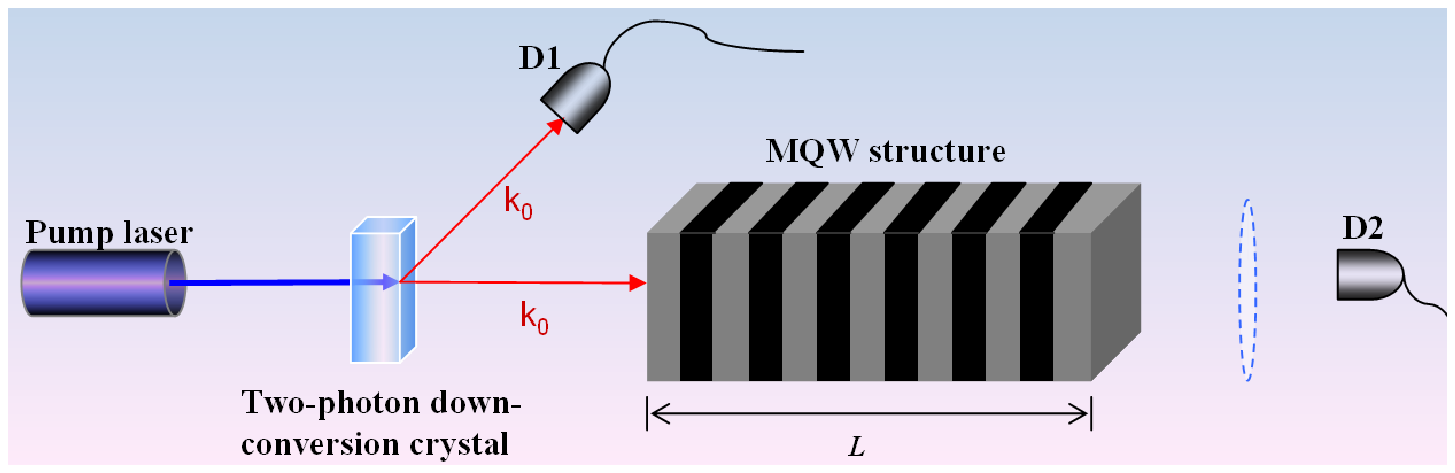,width=14cm}
\end{center}\caption{}
\label{fig1-scheme}
\end{figure}
\clearpage

\begin{figure}
\begin{center}
\epsfig{file=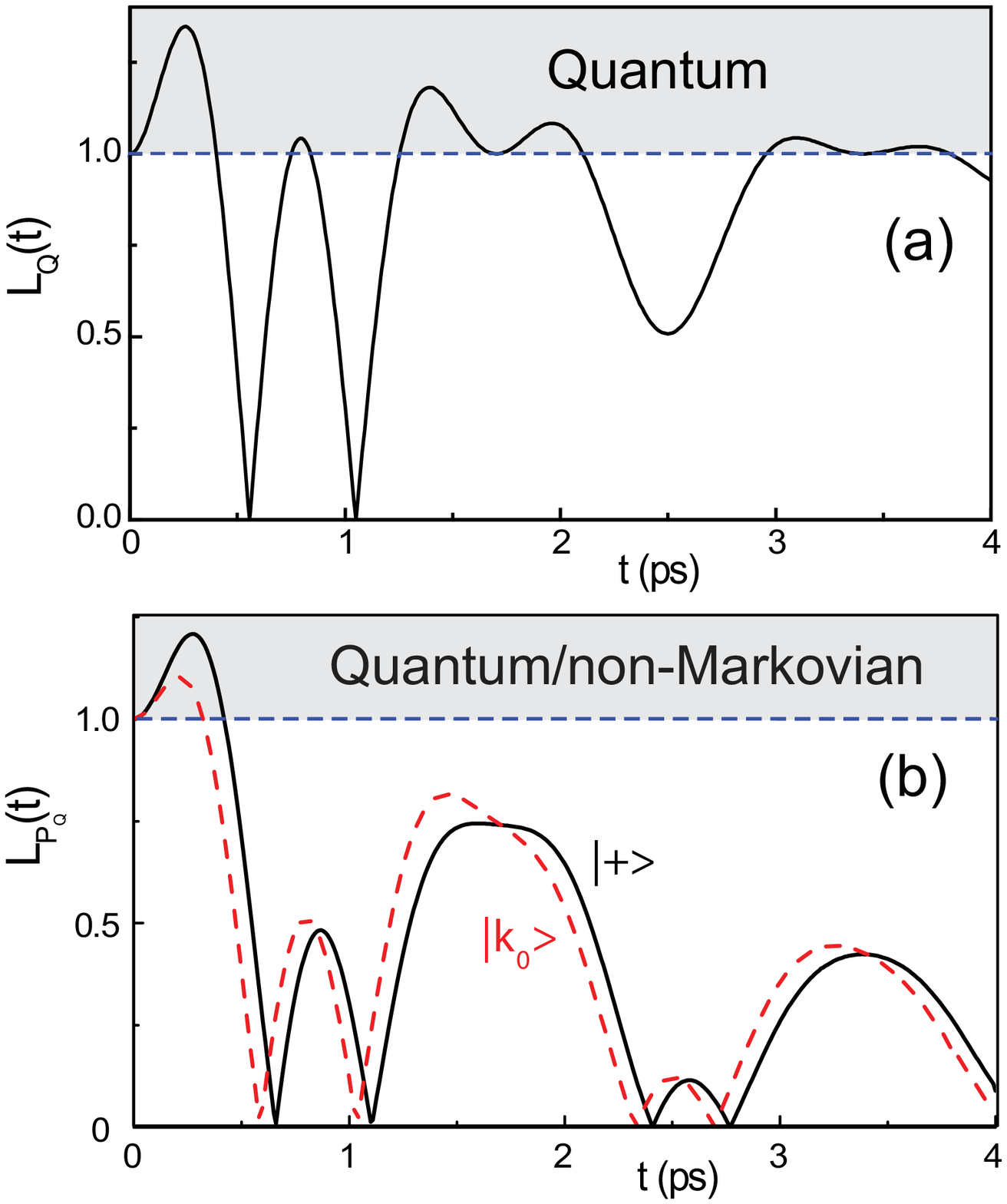,width=14cm}
\end{center}\caption{}
\label{fig1-scheme}
\end{figure}

\clearpage
\begin{figure}
\begin{center}
\epsfig{file=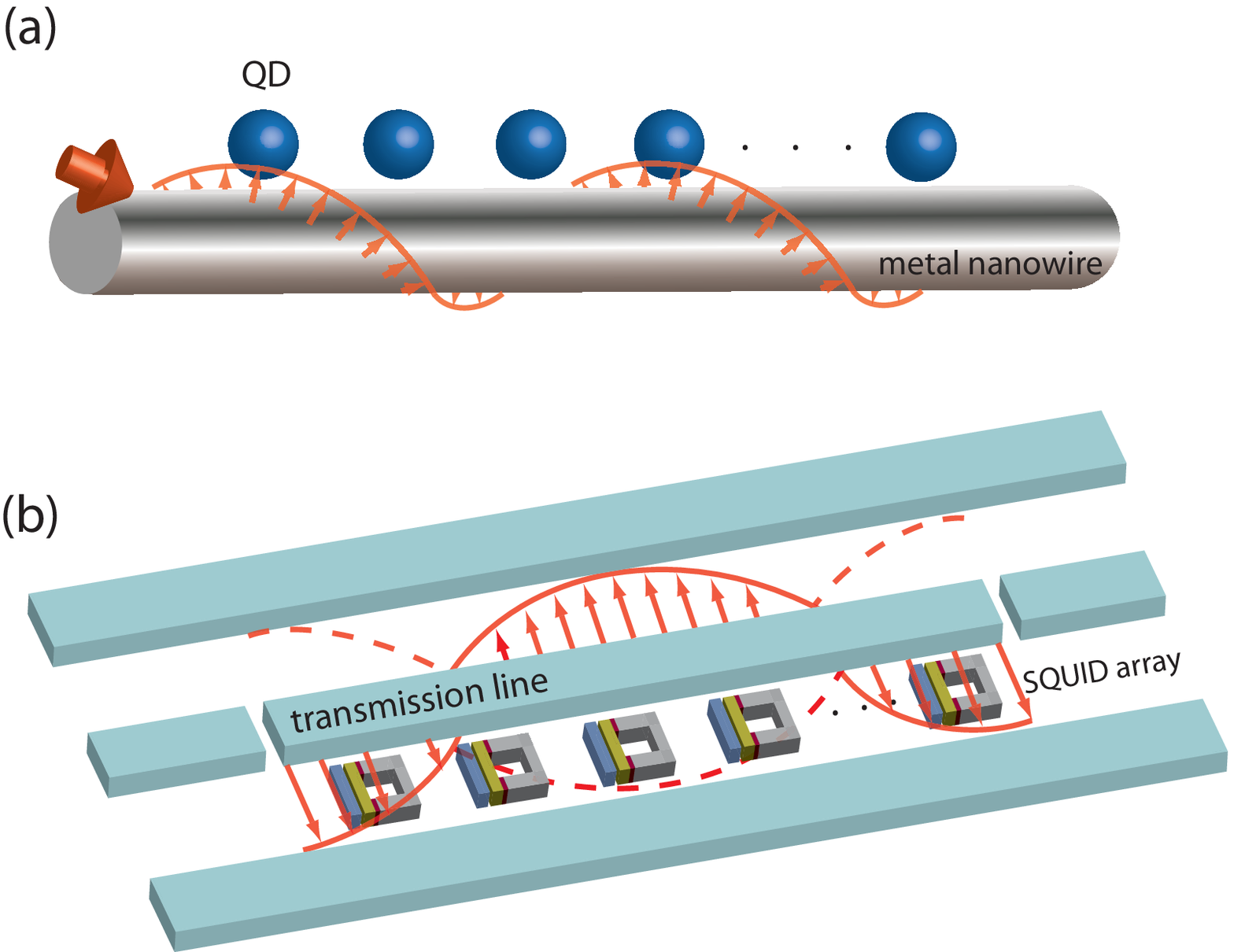,width=14cm}
\end{center}\caption{}
\label{fig1-scheme}
\end{figure}
\clearpage


\clearpage

\begin{thebibliography}{10}
\expandafter\ifx\csname url\endcsname\relax
  \def\url#1{\texttt{#1}}\fi
\expandafter\ifx\csname urlprefix\endcsname\relax\def\urlprefix{URL }\fi
\providecommand{\bibinfo}[2]{#2}
\providecommand{\eprint}[2][]{\url{#2}}

\bibitem{Lvo}
\bibinfo{author}{Lvovsky, A.~I.} \& \bibinfo{author}{Hartmann, S.~R.}
\newblock \bibinfo{title}{Superradiant self-diffraction}.
\newblock \emph{\bibinfo{journal}{Phys. Rev. A}} \textbf{\bibinfo{volume}{59}},
  \bibinfo{pages}{4052--4057} (\bibinfo{year}{1999}).

\bibitem{Inouye}
\bibinfo{author}{Inouye, S.} \emph{et~al.}
\newblock \bibinfo{title}{Superradiant \textsc{R}ayleigh scattering from a
  \textsc{B}ose-\textsc{E}instein condensate}.
\newblock \emph{\bibinfo{journal}{Science}} \textbf{\bibinfo{volume}{285}},
  \bibinfo{pages}{571--574} (\bibinfo{year}{1999}).

\bibitem{SRQD}
\bibinfo{author}{Scheibner, M.} \emph{et~al.}
\newblock \bibinfo{title}{Superradiance of quantum dots}.
\newblock \emph{\bibinfo{journal}{Nat. Phys.}} \textbf{\bibinfo{volume}{3}},
  \bibinfo{pages}{106--110} (\bibinfo{year}{2007}).

\bibitem{SRQW}
\bibinfo{author}{Goldberg, D.} \emph{et~al.}
\newblock \bibinfo{title}{Exciton-lattice polaritons in
  multiple-quantum-well-based photonic crystals}.
\newblock \emph{\bibinfo{journal}{Nat. Photon.}} \textbf{\bibinfo{volume}{3}},
  \bibinfo{pages}{662--666} (\bibinfo{year}{2009}).

\bibitem{zhou}
\bibinfo{author}{Zhou, L.}, \bibinfo{author}{Gong, Z.~R.},
  \bibinfo{author}{Liu, Y.~X.}, \bibinfo{author}{Sun, C.~P.} \&
  \bibinfo{author}{Nori, F.}
\newblock \bibinfo{title}{Controllable scattering of a single photon inside a
  one-dimensional resonator waveguide}.
\newblock \emph{\bibinfo{journal}{Phys. Rev. Lett.}}
  \textbf{\bibinfo{volume}{101}}, \bibinfo{pages}{100501}
  (\bibinfo{year}{2008}).

\bibitem{scullyPRA}
\bibinfo{author}{Svidzinsky, A.}, \bibinfo{author}{Chang, J.~T.} \&
  \bibinfo{author}{Scully, M.~O.}
\newblock \bibinfo{title}{Cooperative spontaneous emission of \textsc{$N$}
  atoms: Many-body eigenstates, the effect of virtual \textsc{L}amb shift
  processes, and analogy with radiation of \textsc{$N$} classical oscillators}.
\newblock \emph{\bibinfo{journal}{Phys. Rev. A}} \textbf{\bibinfo{volume}{81}},
  \bibinfo{pages}{053821} (\bibinfo{year}{2010}).

\bibitem{newPRA}
\bibinfo{author}{Svidzinsky, A.~A.}
\newblock \bibinfo{title}{Nonlocal effects in single-photon superradiance}.
\newblock \emph{\bibinfo{journal}{Phys. Rev. A}} \textbf{\bibinfo{volume}{85}},
  \bibinfo{pages}{013821} (\bibinfo{year}{2012}).

\bibitem{scullyPRL08}
\bibinfo{author}{Svidzinsky, A.}, \bibinfo{author}{Chang, J.~T.} \&
  \bibinfo{author}{Scully, M.~O.}
\newblock \bibinfo{title}{Dynamical evolution of correlated spontaneous
  emission of a single photon from a uniformly excited cloud of \textsc{$N$}
  atoms}.
\newblock \emph{\bibinfo{journal}{Phys. Rev. Lett.}}
  \textbf{\bibinfo{volume}{100}}, \bibinfo{pages}{160504}
  (\bibinfo{year}{2008}).

\bibitem{wstatepaper}
\bibinfo{author}{Wiegner, R.}, \bibinfo{author}{von Zanthier, J.} \&
  \bibinfo{author}{Agarwal, G.~S.}
\newblock \bibinfo{title}{Quantum-interference-initiated superradiant and
  subradiant emission from entangled atoms}.
\newblock \emph{\bibinfo{journal}{Phys. Rev. A}} \textbf{\bibinfo{volume}{84}},
  \bibinfo{pages}{023805} (\bibinfo{year}{2011}).

\bibitem{LG}
\bibinfo{author}{Leggett, A.~J.} \& \bibinfo{author}{Garg, A.}
\newblock \bibinfo{title}{Quantum mechanics versus macroscopic realism: Is the
  flux there when nobody looks?}
\newblock \emph{\bibinfo{journal}{Phys. Rev. Lett.}}
  \textbf{\bibinfo{volume}{54}}, \bibinfo{pages}{857--860}
  (\bibinfo{year}{1985}).

\bibitem{NLPRL}
\bibinfo{author}{Lambert, N.}, \bibinfo{author}{Emary, C.},
  \bibinfo{author}{Chen, Y.~N.} \& \bibinfo{author}{Nori, F.}
\newblock \bibinfo{title}{Distinguishing quantum and classical transport
  through nanostructures}.
\newblock \emph{\bibinfo{journal}{Phys. Rev. Lett.}}
  \textbf{\bibinfo{volume}{105}}, \bibinfo{pages}{176801}
  (\bibinfo{year}{2010}).

\bibitem{scullyscience}
\bibinfo{author}{Scully, M.~O.} \& \bibinfo{author}{Svidzinsky, A.~A.}
\newblock \bibinfo{title}{The super of superradiance}.
\newblock \emph{\bibinfo{journal}{Science}} \textbf{\bibinfo{volume}{325}},
  \bibinfo{pages}{1510} (\bibinfo{year}{2009}).

\bibitem{scullyPRL06}
\bibinfo{author}{Scully, M.~O.}, \bibinfo{author}{Fry, E.~S.},
  \bibinfo{author}{Ooi, C. H.~R.} \& \bibinfo{author}{W\'{o}dkiewicz, K.}
\newblock \bibinfo{title}{Directed spontaneous emission from an extended
  ensemble of \textsc{$N$} atoms: Timing is everything}.
\newblock \emph{\bibinfo{journal}{Phys. Rev. Lett.}}
  \textbf{\bibinfo{volume}{96}}, \bibinfo{pages}{010501}
  (\bibinfo{year}{2006}).

\bibitem{scullylaser}
\bibinfo{author}{Scully, M.~O.}
\newblock \bibinfo{title}{Correlated spontaneous emission on the
  \textsc{V}olga}.
\newblock \emph{\bibinfo{journal}{Laser Phys.}} \textbf{\bibinfo{volume}{17}},
  \bibinfo{pages}{635--646} (\bibinfo{year}{2007}).

\bibitem{Lee}
\bibinfo{author}{Liu, K.~C.} \& \bibinfo{author}{Lee, Y.~C.}
\newblock \bibinfo{title}{Radiative decay of \textsc{W}annier excitons in thin
  crystal films}.
\newblock \emph{\bibinfo{journal}{Physica A}} \textbf{\bibinfo{volume}{102}},
  \bibinfo{pages}{131--144} (\bibinfo{year}{1980}).

\bibitem{YN}
\bibinfo{author}{Chen, Y.~N.} \& \bibinfo{author}{Chuu, D.~S.}
\newblock \bibinfo{title}{Decay rate and renormalized frequency shift of
  superradiant excitons: Crossover from two-dimensional to three-dimensional
  crystals}.
\newblock \emph{\bibinfo{journal}{Phys. Rev. B}} \textbf{\bibinfo{volume}{61}},
  \bibinfo{pages}{10815--10819} (\bibinfo{year}{2000}).

\bibitem{Gurvitz}
\bibinfo{author}{Elattari, B.} \& \bibinfo{author}{Gurvitz, S.~A.}
\newblock \bibinfo{title}{Influence of measurement on the lifetime and the
  linewidth of unstable systems}.
\newblock \emph{\bibinfo{journal}{Phys. Rev. A}} \textbf{\bibinfo{volume}{62}},
  \bibinfo{pages}{032102} (\bibinfo{year}{2000}).

\bibitem{pin}
\bibinfo{author}{Ashkenasy, N.} \emph{et~al.}
\newblock
  \bibinfo{title}{\textsc{G}a\textsc{A}s/\textsc{A}l\textsc{G}a\textsc{A}s
  single quantum well p-i-n structures: A surface photovoltage study}.
\newblock \emph{\bibinfo{journal}{J. Appl. Phys.}}
  \textbf{\bibinfo{volume}{86}}, \bibinfo{pages}{6902--6907}
  (\bibinfo{year}{1999}).

\bibitem{koch}
\bibinfo{author}{Kira, M.}, \bibinfo{author}{Jahnke, F.} \&
  \bibinfo{author}{Koch, S.~W.}
\newblock \bibinfo{title}{Quantum theory of secondary emission in optically
  excited semiconductor quantum wells}.
\newblock \emph{\bibinfo{journal}{Phys. Rev. Lett.}}
  \textbf{\bibinfo{volume}{82}}, \bibinfo{pages}{3544--3547}
  (\bibinfo{year}{1999}).

\bibitem{kavokin}
\bibinfo{author}{Malpuech, G.} \& \bibinfo{author}{Kavokin, A.}
\newblock \bibinfo{title}{Resonant \textsc{R}ayleigh scattering of
  exciton-polaritons in multiple quantum wells}.
\newblock \emph{\bibinfo{journal}{Phys. Rev. Lett.}}
  \textbf{\bibinfo{volume}{85}}, \bibinfo{pages}{650--653}
  (\bibinfo{year}{2000}).

\bibitem{CircuitQED}
\bibinfo{author}{Bishop, L.~S.} \emph{et~al.}
\newblock \bibinfo{title}{Nonlinear response of the vacuum \textsc{R}abi
  resonance}.
\newblock \emph{\bibinfo{journal}{Nat. Phys.}} \textbf{\bibinfo{volume}{5}},
  \bibinfo{pages}{105} (\bibinfo{year}{2009}).

\bibitem{Patton}
\bibinfo{author}{Patton, B.}, \bibinfo{author}{Woggon, U.} \&
  \bibinfo{author}{Langbein, W.}
\newblock \bibinfo{title}{Coherent control and polarization readout of
  individual excitonic states}.
\newblock \emph{\bibinfo{journal}{Phys. Rev. Lett.}}
  \textbf{\bibinfo{volume}{95}}, \bibinfo{pages}{266401}
  (\bibinfo{year}{2005}).

\bibitem{QuantumNoise}
\bibinfo{author}{Gardiner} \& \bibinfo{author}{Zoller, P.}
\newblock \emph{\bibinfo{title}{Quantum Noise}} (\bibinfo{publisher}{Springer,
  Heidelberg}, \bibinfo{year}{2004}).

\bibitem{Laloy}
\bibinfo{author}{Palacios-Laloy, A.} \emph{et~al.}
\newblock \bibinfo{title}{Experimental violation of a \textsc{B}ell's
  inequality in time with weak measurement}.
\newblock \emph{\bibinfo{journal}{Nat. Phys.}} \textbf{\bibinfo{volume}{6}},
  \bibinfo{pages}{442--447} (\bibinfo{year}{2010}).

\bibitem{PNAS}
\bibinfo{author}{Goggin, M.~E.} \emph{et~al.}
\newblock \bibinfo{title}{Violation of the \textsc{L}eggett-\textsc{G}arg
  inequality with weak measurements of photons}.
\newblock \emph{\bibinfo{journal}{Proc. Natl Acad. Sci.}}
  \textbf{\bibinfo{volume}{108}}, \bibinfo{pages}{1256--1261}
  (\bibinfo{year}{2011}).

\bibitem{PRL}
\bibinfo{author}{Dressel, J.}, \bibinfo{author}{Broadbent, C.~J.},
  \bibinfo{author}{Howell, J.~C.} \& \bibinfo{author}{Jordan, A.~N.}
\newblock \bibinfo{title}{Experimental violation of two-party
  \textsc{L}eggett-\textsc{G}arg inequalities with semiweak measurements}.
\newblock \emph{\bibinfo{journal}{Phys. Rev. Lett.}}
  \textbf{\bibinfo{volume}{106}}, \bibinfo{pages}{040402}
  (\bibinfo{year}{2011}).

\bibitem{SR}
\bibinfo{author}{Xu, J.-S.}, \bibinfo{author}{Li, C.-F.}, \bibinfo{author}{Zou,
  X.-B.} \& \bibinfo{author}{Guo, G.-C.}
\newblock \bibinfo{title}{Experimental violation of the
  \textsc{L}eggett-\textsc{G}arg inequality under decoherence}.
\newblock \emph{\bibinfo{journal}{Sci. Rep.}} \textbf{\bibinfo{volume}{1:101}}
  (\bibinfo{year}{2011}).

\bibitem{natcomm}
\bibinfo{author}{Knee, G.~C.} \emph{et~al.}
\newblock \bibinfo{title}{Violation of a \textsc{L}eggett-\textsc{G}arg
  inequality with ideal non-invasive measurements}.
\newblock \emph{\bibinfo{journal}{Nat. Commun.}}
  \textbf{\bibinfo{volume}{3:606}} (\bibinfo{year}{2011}).

\bibitem{PRL_spin}
\bibinfo{author}{Athalye, V.}, \bibinfo{author}{Roy, S.~S.} \&
  \bibinfo{author}{Mahesh, T.~S.}
\newblock \bibinfo{title}{Investigation of the \textsc{L}eggett-\textsc{G}arg
  inequality for precessing nuclear spins}.
\newblock \emph{\bibinfo{journal}{Phys. Rev. Lett.}}
  \textbf{\bibinfo{volume}{107}}, \bibinfo{pages}{130402}
  (\bibinfo{year}{2011}).

\bibitem{PRB_NL}
\bibinfo{author}{Lambert, N.}, \bibinfo{author}{Johansson, R.} \&
  \bibinfo{author}{Nori, F.}
\newblock \bibinfo{title}{Macrorealism inequality for optoelectromechanical
  systems}.
\newblock \emph{\bibinfo{journal}{Phys. Rev. B}} \textbf{\bibinfo{volume}{84}},
  \bibinfo{pages}{245421} (\bibinfo{year}{2011}).

\bibitem{Schulzgen}
\bibinfo{author}{Sch\"{u}lzgen, A.} \emph{et~al.}
\newblock \bibinfo{title}{Direct observation of excitonic \textsc{R}abi
  oscillations in semiconductors}.
\newblock \emph{\bibinfo{journal}{Phys. Rev. Lett.}}
  \textbf{\bibinfo{volume}{82}}, \bibinfo{pages}{2346--2349}
  (\bibinfo{year}{1999}).

\bibitem{Gleyzes}
\bibinfo{author}{Gleyzes, S.} \emph{et~al.}
\newblock \bibinfo{title}{Quantum jumps of light recording the birth and death
  of a photon in a cavity}.
\newblock \emph{\bibinfo{journal}{Nature}} \textbf{\bibinfo{volume}{446}},
  \bibinfo{pages}{297--300} (\bibinfo{year}{2007}).

\bibitem{Khalili}
\bibinfo{author}{Braginsky, V.~B.} \& \bibinfo{author}{Khalili, F.~Y.}
\newblock \bibinfo{title}{Quantum nondemolition measurements: the route from
  toys to tools}.
\newblock \emph{\bibinfo{journal}{Rev. Mod. Phys.}}
  \textbf{\bibinfo{volume}{68}}, \bibinfo{pages}{1--11} (\bibinfo{year}{1996}).

\bibitem{Johnson}
\bibinfo{author}{Johnson, B.~R.} \emph{et~al.}
\newblock \bibinfo{title}{Quantum non-demolition detection of single microwave
  photons in a circuit}.
\newblock \emph{\bibinfo{journal}{Nat. Phys.}} \textbf{\bibinfo{volume}{6}},
  \bibinfo{pages}{663--667} (\bibinfo{year}{2010}).

\bibitem{NLPRA}
\bibinfo{author}{Lambert, N.}, \bibinfo{author}{Chen, Y.~N.} \&
  \bibinfo{author}{Nori, F.}
\newblock \bibinfo{title}{Unified single-photon and single-electron counting
  statistics: From cavity \textsc{QED} to electron transport}.
\newblock \emph{\bibinfo{journal}{Phys. Rev. A}} \textbf{\bibinfo{volume}{82}},
  \bibinfo{pages}{063840} (\bibinfo{year}{2010}).

\bibitem{huelga2}
\bibinfo{author}{Huelga, S.~F.}, \bibinfo{author}{Marshall, T.~W.} \&
  \bibinfo{author}{Santos, E.}
\newblock \bibinfo{title}{Proposed test for realist theories using
  \textsc{R}ydberg atoms coupled to a high-\textsc{$Q$} resonator}.
\newblock \emph{\bibinfo{journal}{Phys. Rev. A}} \textbf{\bibinfo{volume}{52}},
  \bibinfo{pages}{R2497--R2500} (\bibinfo{year}{1995}).

\bibitem{huelga3}
\bibinfo{author}{Huelga, S.~F.}, \bibinfo{author}{Marshall, T.~W.} \&
  \bibinfo{author}{Santos, E.}
\newblock \bibinfo{title}{Temporal \textsc{B}ell-type inequalities for
  two-level \textsc{R}ydberg atoms coupled to a high-\textsc{$Q$} resonator}.
\newblock \emph{\bibinfo{journal}{Phys. Rev. A}} \textbf{\bibinfo{volume}{54}},
  \bibinfo{pages}{1798--1807} (\bibinfo{year}{1996}).

\bibitem{huelga1}
\bibinfo{author}{Waldherr, G.}, \bibinfo{author}{Neumann, P.},
  \bibinfo{author}{Huelga, S.~F.}, \bibinfo{author}{Jelezko, F.} \&
  \bibinfo{author}{Wrachtrup, J.}
\newblock \bibinfo{title}{Violation of a temporal \textsc{B}ell inequality for
  single spins in a diamond defect center}.
\newblock \emph{\bibinfo{journal}{Phys. Rev. Lett.}}
  \textbf{\bibinfo{volume}{107}}, \bibinfo{pages}{090401}
  (\bibinfo{year}{2011}).

\bibitem{PRB_GY}
\bibinfo{author}{Chen, G.-Y.}, \bibinfo{author}{Lambert, N.},
  \bibinfo{author}{Chou, C.-H.}, \bibinfo{author}{Chen, Y.-N.} \&
  \bibinfo{author}{Nori, F.}
\newblock \bibinfo{title}{Surface plasmons in a metal nanowire coupled to
  colloidal quantum dots: Scattering properties and quantum entanglement}.
\newblock \emph{\bibinfo{journal}{Phys. Rev. B}} \textbf{\bibinfo{volume}{84}},
  \bibinfo{pages}{045310} (\bibinfo{year}{2011}).

\bibitem{Mpho1}
\bibinfo{author}{Romero, G.}, \bibinfo{author}{Garc\'{i}a-Ripoll, J.~J.} \&
  \bibinfo{author}{Solano, E.}
\newblock \bibinfo{title}{Microwave photon detector in circuit \textsc{QED}}.
\newblock \emph{\bibinfo{journal}{Phys. Rev. Lett.}}
  \textbf{\bibinfo{volume}{102}}, \bibinfo{pages}{173602}
  (\bibinfo{year}{2009}).

\bibitem{Mpho2}
\bibinfo{author}{Peropadre, B.} \emph{et~al.}
\newblock \bibinfo{title}{Approaching perfect microwave photodetection in
  circuit \textsc{QED}}.
\newblock \emph{\bibinfo{journal}{Phys. Rev. A}} \textbf{\bibinfo{volume}{84}},
  \bibinfo{pages}{063834} (\bibinfo{year}{2011}).

\bibitem{Mpho3}
\bibinfo{author}{Chen, Y.-F.} \emph{et~al.}
\newblock \bibinfo{title}{Microwave photon counter based on \textsc{J}osephson
  junctions}.
\newblock \emph{\bibinfo{journal}{Phys. Rev. Lett.}}
  \textbf{\bibinfo{volume}{107}}, \bibinfo{pages}{217401}
  (\bibinfo{year}{2011}).

\bibitem{Mpho4}
\bibinfo{author}{Filipp, S.} \emph{et~al.}
\newblock \bibinfo{title}{Two-qubit state tomography using a joint dispersive
  readout}.
\newblock \emph{\bibinfo{journal}{Phys. Rev. Lett.}}
  \textbf{\bibinfo{volume}{102}}, \bibinfo{pages}{200402}
  (\bibinfo{year}{2009}).

\bibitem{Mpho5}
\bibinfo{author}{Reed, M.~D.} \emph{et~al.}
\newblock \bibinfo{title}{Realization of three-qubit quantum error correction
  with superconducting circuits}.
\newblock \emph{\bibinfo{journal}{Nature}} \textbf{\bibinfo{volume}{482}},
  \bibinfo{pages}{382--385} (\bibinfo{year}{2012}).

\end{thebibliography}
\end{document}